\documentclass[aps,pra,twocolumn,showpacs,amsmath,amssymb,superscriptaddress,longbibliography]{revtex4-2}
\usepackage[utf8]{inputenc}
\usepackage{graphicx,graphics}
\usepackage[title,titletoc]{appendix}
\usepackage{hyperref}
\hypersetup{colorlinks=true,linkcolor=blue,urlcolor=blue,citecolor=blue}
\begin{document}
\title{Mediated Interactions and Damping Effects in Superfluid Mixtures of Bose and Fermi Gases}

\author{Dong-Chen Zheng}
\affiliation{Fujian Provincial Key Laboratory for Quantum Manipulation and New Energy Materials, College of Physics and Energy, Fujian Normal University, Fuzhou 350117, China}
\affiliation{Fujian Provincial Collaborative Innovation Center for Advanced High-Field Superconducting Materials and Engineering, Fuzhou, 350117, China}
\author{Yu-Xin Liao}
\affiliation{Fujian Provincial Key Laboratory for Quantum Manipulation and New Energy Materials, College of Physics and Energy, Fujian Normal University, Fuzhou 350117, China}
\affiliation{Fujian Provincial Collaborative Innovation Center for Advanced High-Field Superconducting Materials and Engineering, Fuzhou, 350117, China}
\author{Wen Lin}
\affiliation{College of Physics and Electronic Engineering, Chongqing Normal University, Chongqing 401331, China}
\author{Renyuan Liao}\email{ryliao@fjnu.edu.cn}
\affiliation{Fujian Provincial Key Laboratory for Quantum Manipulation and New Energy Materials, College of Physics and Energy, Fujian Normal University, Fuzhou 350117, China}
\affiliation{Fujian Provincial Collaborative Innovation Center for Advanced High-Field Superconducting Materials and Engineering, Fuzhou, 350117, China}
\date{\today}
\begin{abstract}
  We investigate the homogeneous superfluid mixtures of Bardeen-Cooper-Schrieffer~(BCS) superfluid originating from pairing two-species fermionic atoms and superfluidity stemming from condensation of bosonic atoms. By integrating out the freedoms associated with the BCS superfluid, we derive the fermion-mediated interactions between bosons, which is attractive and can be tuned from long range in the BCS region to short range in the region of Bose-Einstein condensation (BEC) of molecular dimers. By analyzing the Bogoliubov spectrum and the damping rate of bosonic superfluid, we map out the phase diagram spanned by the boson-fermion mass ratio and the boson-fermion coupling strength, which consists of a phase separation region and two phase mixing regions with and without Landau damping. The three different phases can coexist at a tricritical point, which moves toward low boson-fermion mass ratio and high boson-fermion scattering length as the fermion-fermion interaction strength is tuned up on the BCS side.
\end{abstract}
\maketitle
\section{Introduction}
Mediated interactions play a crucial role in our understanding of nature. In particle physics, all fundamental interactions are mediated by gauge bosons~\cite{WEI95}. In condensed matter physics, phonon-mediated electron-electron attractions are responsible for the formation of Cooper pairs, whose condensation leads to the phenomena of conventional superconductivity~\cite{GIR19}. Ultracold atoms have emerged as an ideal platform for engineering the interatomic interactions~\cite{CHI10,RIT13,DAL11}, testing the fundamental physics~\cite{GEO14} and exploring the novel many-body quantum phenomena~\cite{BLO08}. Of particular interests are the experimental observations of the fermion-mediated long-range interactions between bosons in Bose-Fermi mixtures in weakly interacting~\cite{CHI19,BRU19,EDR20} and strongly interacting regimes~\cite{CHI23,CHE22}. This has sparked new interests in theoretical investigating of physics associated with the fermion-mediated interactions in various physical systems. These includes studying the stability conditions for weakly-interacting Bose-Fermi mixtures at zero temperature~\cite{LIA21,LRY20} and at finite temperatures~\cite{OHA21}, investigating mediated interactions with strong coupling theories~\cite{BRU22,JIA21} and effective field theories~\cite{TIL22}, and tailoring long-range interactions for quantum simulators~\cite{DAN22}.

Superfluid mixtures of bosonic and fermionic atoms have been the focus of both theoretical~\cite{PU11,OZA14,HUI14,ZHA14,BRU15,DAL16,ZHA17,SOR19,KAW20} and experimental~\cite{SAL14,SAL15,PAN16,GUP17,PAN18} researches over the past years. These double superfluid systems provide fascinating opportunities to explore the interplay between excitations of distinct statistics and mediated interactions. A Bose-Fermi superfluid mixture possesses two gapless bosonic modes resulting from the spontaneous breaking of internal gauge symmetries of Bose superfluid and Fermi superfluid, respectively, and a gapped fermionic excitations that describes the Cooper pair breaking~\cite{ZHA14,HOI17,KUR23}. One of the key questions to ask is how fermion-mediated interactions reshape our understanding of this exciting system. While existing experiments~\cite{SAL14,PAN18} on double superfluid mixtures indicates damping of dipole modes, searching for well-defined quasiparticle excitations in interacting quantum matter represents one of the cornerstones of modern physics~\cite{ZWI20}. Superfluid mixtures of Bose-Fermi gases offer promising prospects to elucidate the physics of such quasiparticles.

 In this work, we are trying to address this question by conducting the following studies: First, we will start from the functional integral representation of the partition function of the system. By tracing out the fermions, we obtain an effective action entirely in terms of  degrees of freedom associated with bosons, so that we can isolate the effects of fermion-mediated interactions on the bosons. Second, we will examine how the induced interactions modify the Bogoliugov spectrum of bosons and lead to the damping of quasiparticles. Third, we will map out the phase diagrams emphasizing the roles of boson-fermion mass ratio and boson-fermion interaction strength. Finally, by determining the behaviors of the tricritical point as a function of inter-fermion scattering length, we can completely characterize the topology of the phase diagram without recourse to extensive numerical treatment.
\section{Model and Formalism}
We consider a homogeneous mixture of bosons and population balanced spin-1/2 fermions, described by the following grand canonical Hamiltonian:
\begin{eqnarray}
H &=&\int d^{3}\mathbf{r}\bigg[\sum_{\sigma =\uparrow ,\downarrow }\psi _{\sigma }^{\dag }\left(h_F +g_{BF}\phi ^{\dag }\phi \right) \psi
_{\sigma }+g_{F}\psi _{\uparrow }^{\dag }\psi _{\downarrow }^{\dag }\psi
_{\downarrow }\psi _{\uparrow }\notag \\
&&+\phi ^{\dag }h_B \phi +\frac{g_{B}}{2}\phi ^{\dag }\phi ^{\dag
}\phi \phi\bigg],
\end{eqnarray}
where $h_{i}=-\frac{\hbar^2}{2m_{i}}\nabla^2-\mu_{i}$, $i=B,F$ denotes bosons or fermions with mass $m_i$, and $\mu_i$ represents the chemical potential. $\phi$ and $\psi_{\sigma}$ are the field operators for bosons and fermions with spin $\sigma=\uparrow,\downarrow$, respectively. In bose gases, $g_B=4\pi\hbar^2a_B/m_B$ with positive $s$-wave scattering length $a_B$ characterizes the repulsive interaction strength between bosons. In fermi gases, $g_F$ is the interaction strength between fermions and assumed to be attractive, leading to BCS pairing. $g_{BF}=2\pi\hbar^2a_{BF}(m_F^{-1}+m_B^{-1})$ accounts for the interaction strength between fermions and bosons, with $a_{BF}$ being the corresponding $s$-wave scattering length. For convenience, we define the Fermi momentum $k_F=(3\pi^2n_F)^{1/3}$ with $n_F$ being the number density of Fermi gases, the Fermi velocity $v_F=\hbar k_F/m_F$ and the corresponding Fermi energy $E_F=\hbar^2k_F^2/2m_F$. We will adopt the natural units by setting $\hbar=k_B=1$ for sake of simplicity from now on.

Within the framework of imaginary-time field integral~\cite{SIM10}, we can cast the partition function of the system as $\mathcal{Z}=\int \mathcal{D}[\bar{\psi}_\sigma,\psi_\sigma]\mathcal{D}[\phi^*,\phi]e^{-S}$, with the action given by $S=\int_0^\beta d\tau \left[H+\int d^3\mathbf{r}\left(\sum_\sigma \bar{\psi}_\sigma\partial_\tau\psi_\sigma+\phi^*\partial_\tau\phi\right)\right]$, where $\beta=1/T$ is the inverse temperature. By performing a Hubbard-Stratonovich transformation, we introduce a bosonic field $\Delta(\mathbf{r},\tau)$, which serves as an order parameter~\cite{NOTEOP} encapsulating the relevant low-energy degrees of freedom for fermions. After carrying out the functional integration over the Grassmann fields, we can obtain an effective action $S_{\text{eff}}=\int d\tau d^3\mathbf{r}\left[\phi^*\left(\partial_\tau+h_B+\frac{g_{B}}{2}\phi^\ast\phi\right)\phi-|\Delta|^2/g_F\right]-\text{Tr}\ln{\mathcal{M}}+\text{Tr}\hat{h}$ with $\hat{h}=-\nabla^2/2m_F-\mu_F+g_{BF}\phi^*\phi$, where the matrix $\mathcal{M}$ reads
\begin{eqnarray}
    \mathcal{M}=\begin{pmatrix}\partial_\tau+\hat{h} & -\Delta\\-\Delta^* &\partial_\tau-\hat{h}
    \end{pmatrix}.
\end{eqnarray}
So far, the above formal manipulation of the partition function is exact.

To facilitate the evaluation of the traces by benefiting the translational invariance, we will transform the above to momentum-frequency representation $\left[q\equiv\left(\mathbf{q},\omega_n\right)\right]$. By making the Fourier expansions $\Delta=\Delta_0+\sum_{q\neq 0}\Delta_qe^{iqx}$ (we shall set $\Delta_0$ to be real) and $\phi^*\phi=\rho_0+\sum_{q\neq 0}\rho_qe^{iqx}$ with space-time coordinate $x=(\mathbf{r},\tau)$, and defining the inverse Green's function $\mathcal{G}^{-1}=-\partial_\tau+(\nabla^2/2m_F+\mu_F-g_{BF}\rho_0)\sigma_z+\Delta_0\sigma_x$ with $\sigma_{x}$ and $\sigma_z$ being the Pauli matrices, we can write $\mathcal{M}=-\mathcal{G}^{-1}+M_1$, where the matrix $M_1$ is
\begin{eqnarray}
      M_1=\sum_{q\neq 0}e^{iqx}\begin{pmatrix}g_{BF}\rho_q & -\Delta_q  \\-\Delta_q^* & g_{BF}\rho_q
     \end{pmatrix}.
\end{eqnarray}
 This allows one to write $\text{Tr}\ln\mathcal{M}=\text{Tr}\ln(-\mathcal{G}^{-1})+\text{Tr}\ln{(I-\mathcal{G}M_1)}$ with the unit matrix $I$, and to perform the series expansion $-\text{Tr}\ln(I-\mathcal{G}M_1))=\sum_{l}\text{Tr}[(\mathcal{G}M_1)^l]/l$ with positive integer $l$. For $l=1$ and $l=2$, we have
\begin{subequations}
\begin{eqnarray}
   \text{Tr}\left(\mathcal{G}M_1\right)&=&M_1(0)\sum_k \mathcal{G}(k)=0,\\
   \text{Tr}\left(\mathcal{G}M_1\right)^2&=&\sum_{kq}\mathcal{G}(k)M_1(-q)\mathcal{G}(k+q)M_1(q).
\end{eqnarray}
\end{subequations}
For $l\geq3$, the related terms are usually related to the induced three-body or more than three-body interactions for bosons, which can be neglected for the dilute gases considered in this work. Therefore, up to the level of Random-Phase-Approximation (RPA), the effective action contains up to the quadratic order of the fluctuating fields $\Delta_q^*$ and $\Delta_q$, which can be integrated out to yield an approximate effective action solely in terms of fields of Bose gases:
\begin{eqnarray}
     &&S_{\text{eff}}=\!\!\int\!\! d\tau d^3\mathbf{r}\phi^*\!\!\left(\partial_\tau\!+\!h_B\!+\!\frac{g_B}{2}\phi^\ast\phi\right)\!\phi\!+\!\frac{g_{BF}^2}{2}\!\sum_{q\neq 0}\Pi(q)\rho_{-q}\rho_q\nonumber\\
     &+&\!\beta V\!\left(\sum_\mathbf{k}\xi_\mathbf{k}-\frac{|\Delta_0|^2}{g_F}\right)\!-\!\text{Tr}\ln\left(-\mathcal{G}^{-1}\right)+\text{Tr}\ln\Gamma^{-1}(q)\label{eq:5a},
\end{eqnarray}
where $V$ represents the volume occupied by the system, $\text{Tr}\ln\Gamma^{-1}(\mathbf{q},z)$ is the so-called Nozie$\acute{r}$es-Schmitt-Rink (NSR) correction~\cite{NSR85,RAN93} and $\Pi(\mathbf{q},z)=\Pi_{pb}(\mathbf{q},z)+\Pi_{cl}(\mathbf{q},z)$ with $\Pi_{cl}=-|\Delta_0|^2\left(I_{11}A^2+z^2I_{22}B^2-2z^2I_{12}AB\right)/\left(I_{11}I_{22}-z^2I_{12}^2\right)$ and $z=i\omega_n$ is the polarization function, describing the response of the superfluid Fermi gases under external density perturbation, $\omega_n$ is the bosonic Matsubara frequency. And $\Gamma^{-1}(q)$ and the parameters in the polarization function are given as follows
\begin{subequations}
\begin{eqnarray}
   A(\mathbf{q},z)\!&=&\!\sum_\mathbf{p}\!\frac{E_+\!+\!E_-}{E_+E_-}\frac{\xi_+\!+\!\xi_-}{z^2\!-\!(E_+\!+\!E_-)^2},\\
   B(\mathbf{q},z)\!&=&\!\sum_\mathbf{p}\!\frac{E_+\!+\!E_-}{E_+E_-}\frac{1}{z^2\!-\!(E_+\!+\!E_-)^2},\\
   \Pi_{pb}(\mathbf{q},z)\!&=&\!\sum_\mathbf{p}\!\frac{E_+\!+\!E_-}{E_+E_-}\frac{E_+E_-\!-\!\xi_+\xi_-\!+\!|\Delta_0|^2}{z^2\!-\!(E_+\!+\!E_-)^2},\\
     I_{11}(\mathbf{q},z)\!&=&\!\sum_\mathbf{p}\!\frac{E_+\!+\!E_-}{E_+E_-}\frac{E_+E_-\!+\!\xi_+\xi_-\!+\!|\Delta_0|^2}{z^2\!-\!(E_+\!+\!E_-)^2}\!+\!\frac{1}{E_\mathbf{p}},\quad\quad\\
  I_{22}(\mathbf{q},z)\!&=&\!\sum_\mathbf{p}\!\frac{E_+\!+\!E_-}{E_+E_-}\frac{E_+E_-\!+\!\xi_+\xi_-\!-\!|\Delta_0|^2}{z^2\!-\!(E_+\!+\!E_-)^2}\!+\!\frac{1}{E_\mathbf{p}},\quad\quad\\
  I_{12}(\mathbf{q},z)\!&=&\!\sum_\mathbf{p}\!\frac{1}{E_+E_-}\frac{E_+\xi_-\!+\!E_-\xi_+}{z^2\!-\!(E_+\!+\!E_-)^2},\\
  \Gamma^{-1}(\mathbf{q},z)&=&-\frac{1}{g_F}+\sum_\mathbf{p}\frac{1-n_F(\xi_\mathbf{p})-n_F(\xi_\mathbf{p+q})}{z-\xi_\mathbf{p}-\xi_\mathbf{p+q}},
\end{eqnarray}
\end{subequations}
where $\pm$ is a shorthand notation for momentum $\mathbf{p}\pm\mathbf{q}/2$, $\xi_\mathbf{p}=\mathbf{p}^2/2m_F-\mu_F+g_{BF}|\phi_0|^2$ and $E_\mathbf{p}=\sqrt{\xi_\mathbf{p}^2+|\Delta_0|^2}$. It is interesting to notice that our approach recovers the same form of density-density correlation function $\Pi(\mathbf{q},\omega+i0^\dagger)$ obtained in studying collective modes with dynamical BCS model formulated with a diagrammatic approach~\cite{STR06} and in studying dissipation of a moving impurity with time-dependent Bogoliugov-deGennes equations~\cite{LRY19} in superfluid Fermi gases. It involves two contributions, one is from the pair-breaking excitations and the other is from the collective excitations.

We perform the standard Bogoliugov decomposition by writing $\phi=\phi_0+\varphi$, where $\phi_0$ and $\varphi$ are the mean-field and fluctuating parts of the bosonic field, respectively. By retaining the fluctuating fields up to quadratic order, we approximate the effective action as $S_{\text{eff}}=S_0+S_g$, where $S_0$ is the mean-field action and $S_g$ is the Gaussian action containing the quadratic orders of $\varphi$ and $\varphi^*$. Employing $\Omega=-\ln{\mathcal{Z}}/\beta V$, we obtain the grand potential density of the system at mean-field level as
\begin{eqnarray}
 \Omega^{(0)}&=&-\mu_B|\phi_0|^2+\frac{g_B}{2}|\phi_0|^4-\frac{|\Delta_0|^2}{g_F}\nonumber\\
 &&+\frac{1}{V}\sum_\mathbf{k}(\xi_\mathbf{k}-E_\mathbf{k})-\frac{2}{\beta V}\sum_\mathbf{k}\ln{\left(1+e^{-\beta E_\mathbf{k}}\right)}.\quad\quad
\end{eqnarray}
In the above, the NSR correction term has been dropped, since the crucial element of the gaussian fluctuation theory is that the relation between the order parameter and the chemical potential is determined by the extremum of the mean-field grand potential $\Omega^{(0)}$ rather than the full grand potential~\cite{DIE08,HE16}. Minimization of $\Omega^{(0)}$ with respect to $\Delta_0^*$ gives the gap equation $-1/g_F=(1/V)\sum_\mathbf{k}\tanh(\beta E_\mathbf{k}/2)/(2E_\mathbf{k})$. Thermodynamic relation $n_F=-\partial\Omega^{(0)}/\partial \mu_F$ gives the number equation $n_F=(1/V)\sum_\mathbf{k}\left[1-\tanh(\beta E_\mathbf{k}/2)\xi_\mathbf{k}/E_\mathbf{k}\right]$. These two equations determine the order parameter $\Delta_0=\Delta_0^c$ and chemical potential $\mu_F=\mu_F^c+g_{BF}n_B$ self-consistently, where $\Delta_0^c$ and $\mu_F^c$ are the solutions in absence of coupling with bosons, and $n_B$ is the number
density of Bose gases. Saddle point condition $\partial\Omega^{(0)}/\partial\phi_0^*=0$ leads to the Hugenholz-Pines theorem~\cite{HP59}, yielding the relation $\mu_B=g_Bn_B+g_{BF}n_F$. At zero temperature, the corresponding ground state energy density  is found from the relation $E_G^{(0)}=\Omega^{(0)}+\mu_F n_F+\mu_B n_B$ yielding
\begin{subequations}
\begin{eqnarray}
 E_G^{(0)}&=&\alpha n_F E_F+\frac{g_B}{2}n_B^2+g_{BF}n_Fn_B,\\
 \alpha(\eta)&=&\frac{\mu_F^c}{E_F}-\frac{3\pi}{8k_Fa}\frac{|\Delta_0|^2}{E_F^2}\nonumber\\
 &&+\frac{1}{n_FE_FV}\sum_\mathbf{k}\left(\xi_\mathbf{k}+\frac{|\Delta_0|^2}{2\epsilon_\mathbf{k}}-E_\mathbf{k}\right).
\end{eqnarray}
\end{subequations}
In the above we have expressed the bare coupling parameter $g_F$ in favor of physical scattering length $a$ via the prescription $1/g_F=m_F/(4\pi a)-(1/V)\sum_\mathbf{k}1/(2\epsilon_\mathbf{k})$ with $\epsilon_\mathbf{k}=\mathbf{k}^2/(2m_F)$. As seen above, the dimensionless coefficient $\alpha(\eta)$ is fully determined by the coupling parameter $\eta\equiv1/(k_Fa)$. Typically, in the deep BCS limit, we have $\alpha$ approaching $3/5$, recovering the well-known result for free fermions~\cite{VIV00,LIA21}.

To ensure the stability of the system, we require that the Hessian matrix $\partial^2 E_G^{(0)}/\partial n_i\partial n_j$ with $i,j=F,B$ constructed for the ground state to be positive definite, which leads to an upper bound for fermion density
\begin{eqnarray}
  n_F^{1/3}<\frac{5(3\pi^2)^{2/3}g_B}{9m_F g_{BF}^2}\left[\alpha-\frac{3}{5}\eta\frac{\partial\alpha}{\partial\eta}+\frac{1}{10}\eta^2\frac{\partial^2\alpha}{\partial\eta^2}\right],
\end{eqnarray}
which is a generalization of the mechanical stability condition for Bose-Fermi mixtures~\cite{VIV00}. Noticing that the sound velocity of the BCS system can be determined via $v_s^{2}=(n_F/m_F)\partial^2E_G^{(0)}/\partial n_F^2$, we obtain an equivalent stability condition for the system against phase separation as $n_F<v_s^2m_Fg_B/g_{BF}^2$, which has been checked consistently in numerics.

\section{Results and Discussion}
\begin{figure}[t]
\includegraphics[width=8cm]{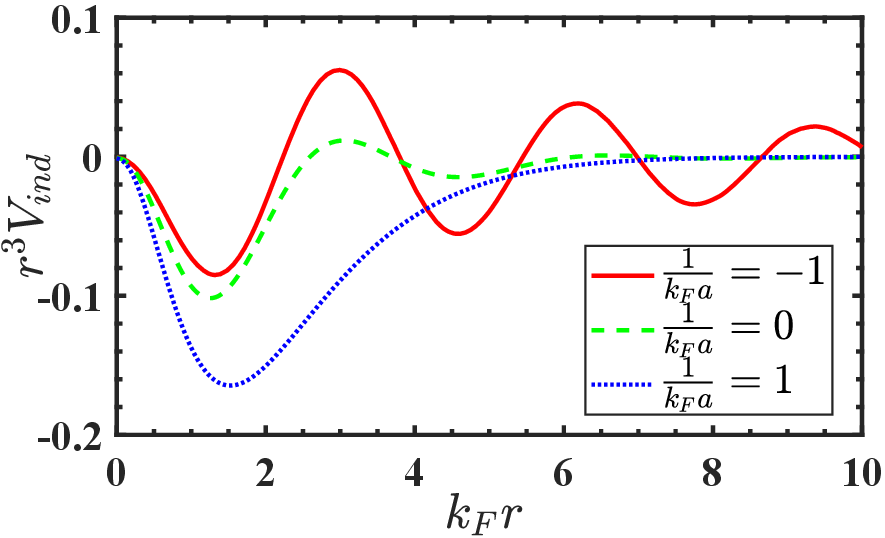}
\caption{(color online) The spatial distribution of $r^3$ scaling of the induced interaction potential $r^3V_{ind}(r)$ [in units of $g_{BF}^2d(E_F)$] between two bosons with relative coordinate $\mathbf{r}$ for three typical interaction parameters $k_Fa=-1$, 0 and 1, corresponding respectively to the regions of BCS, unitarity, and BEC. $d(E_F)=m_Fk_F/\pi^2$ is the density of states of free Fermi gases at the Fermi energy.}
\label{fig1}
\end{figure}

Inspecting the effective action in Eq.~$(\ref{eq:5a})$ and the polarization function, one can obtain the Hamiltonian describing the induced two-body interactions between bosons through coupling with fermions, that is $H_{ind}=(g_{BF}^2/2)\sum_{\mathbf{q}\neq 0}\sum_\mathbf{k,p}\Pi_\mathbf{q}\phi_{\mathbf{k+q}}^\dagger\phi_{\mathbf{p-q}}^\dagger\phi_\mathbf{p}\phi_\mathbf{k}$, where $\Pi_\mathbf{q}\equiv \Pi(\mathbf{q},0)$ is the polarization function evaluated at the static limit at zero temperature. Correspondingly, an induced pairwise interaction potential between two Bose atoms with relative coordinate $\mathbf{r}$ is given by $V_{ind}(\mathbf{r})=\sum_\mathbf{q\neq 0}g_{BF}^2\Pi_\mathbf{q}e^{i\mathbf{q}\cdot \mathbf{r}}$. The $r^3$ scaling behavior of the induced potential $r^3V_{ind}$ is presented in Fig.~\ref{fig1}. The essential features of the fermion-mediated interaction potential are remarkable: On the BCS side with $1/k_Fa=-1$, the potential shows an oscillating power-law behavior ($1/r^3$), a signature of the oscillatory Ruderman-Kittel-Kasuya-Yosida (RKKY) type interaction~\cite{RKKY54}. The RKKY interaction originally describes the effective interaction between two localized magnetic impurities due to the polarization of the conduction electrons near the Fermi surface. For Bose-Fermi mixtures, the effective interaction potential between bosons mediated by  a single species fermions are predicted~\cite{SAN08,SPI14,BRU17} to be of RKKY type in real space, where it decays at $1/r^3$ at large spatial separation and shows the Friedel oscillations at a period of $1/2k_F$, imprinted by the density of the Fermi gases. At unitary with $1/k_Fa=0$, the potential still shows power-law behavior, but with small range. On the BEC side with $1/k_Fa=1$, the potential decreases to zero rather quickly as expected, as at the BEC  limit it has been shown to induce an attractive Yukawa potential (~$e^{-\sqrt{2}r/\xi}/r$) that falls off exponentially beyond the healing length $\xi$~\cite{NAK16,BRU18,BRU18_2}.

\begin{figure}[t]
\includegraphics[width=8cm]{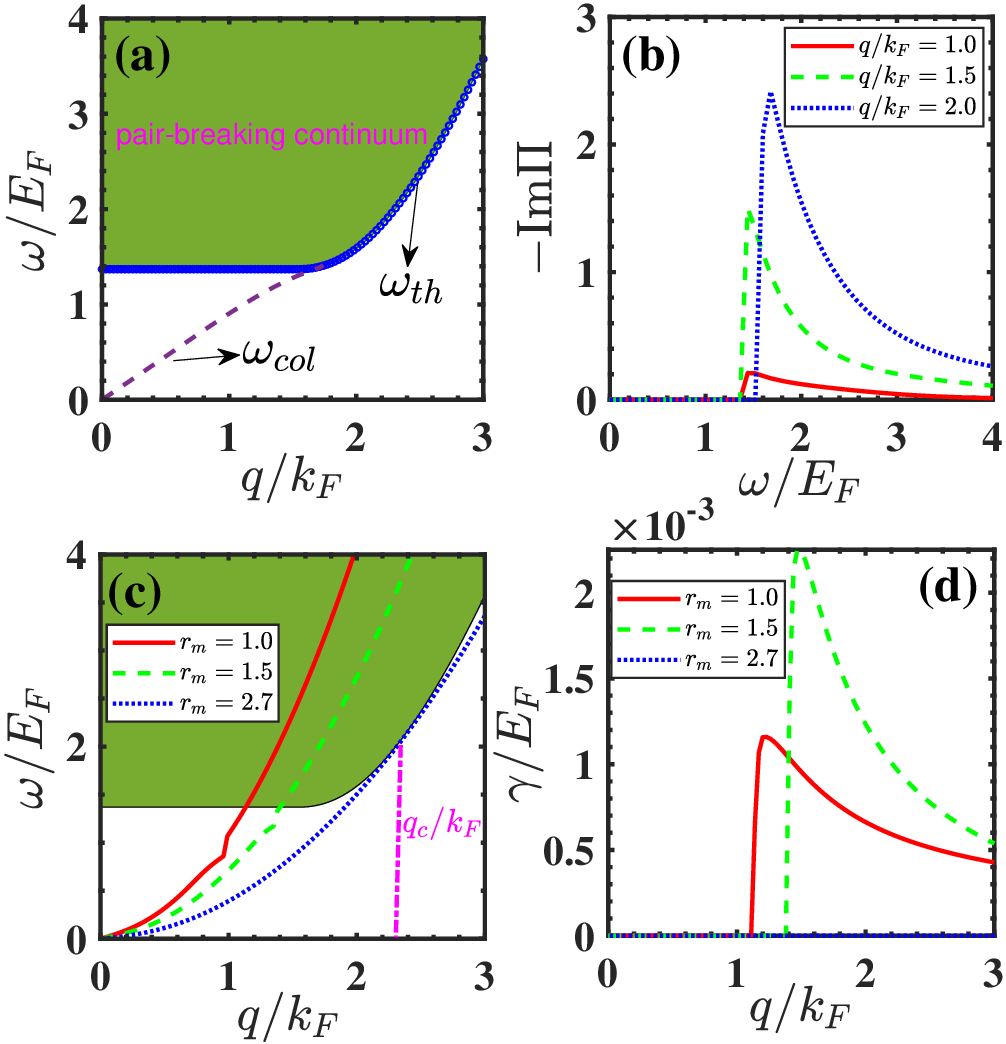}
\caption{(color online) Physics of excitations and damping at the unitarity where $1/k_Fa=0$. Shown in the upper panel are properties of the polarization function: (a)The shaded region is the range where the imaginary part of the polarization differs from zero, and is referred as pair-breaking continuum. $\omega_{\text{th}}$ denotes the threshold for pair-breaking excitation, and $\omega_{\text{col}}$ is the collective excitation. (b)The imaginary part of the polarization function $\text{Im}\Pi(\mathbf{q},\omega)$ [in units of $d(E_F)$] as a function of frequency $\omega$ for given typical momentum amplitudes $q$. Shown in the lower panel are properties of the Bogoliubov quasiparticles: (c)the excitation energy $\omega/E_F$ and (d) the Landau damping rate $\gamma/E_F$ for three typical mass ratio $r_m=1.0$, $1.5$ and $2.7$. $d(E_F)=m_Fk_F/\pi^2$ is the density of states of free Fermi gases at the Fermi energy. The relevant parameters chosen here are $k_Fa_{B}=0.1$, $k_Fa_{BF}=0.05$ and $n_B/n_F=1$.}
\label{fig2}
\end{figure}

The gaussian action for the bosonic fluctuating fields can be compactly written as $S_g=\frac{1}{2}\sum_q\Phi_q^\dagger\mathcal{G}_B^{-1}\Phi_q$ by defining a column vector $\Phi_q=(\varphi_q,\varphi_{-q}^*)^T$ and an inverse matrix $\mathcal{G}_B^{-1}=\epsilon_q+A_q-i\omega_n\sigma_z+A_q\sigma_x$ with $\epsilon_\mathbf{q}=\mathbf{q}^2/2m_B$ and $A_q=(g_{BB}+g_{BF}^2\Pi_q)n_B$. The quasiparticle spectrum $\omega(\mathbf{q})$ and the damping rate $\gamma(\mathbf{q})$ can be obtained by seeking solutions of the secular equation $det\mathcal{G}_B^{-1}(\mathbf{q},\omega-i\gamma)=0$ with substitution of $\Pi_q|{i\omega_n\rightarrow \omega+i0^\dagger}$. By analytic continuation to real frequency~($i\omega_n\rightarrow\omega+i0^\dagger$), one obtains the polarization function $\Pi(\mathbf{q},\omega)$, whose imaginary part provides an essential information for the damping of the excitations of Bose gases. The imaginary part of the polarization is closely related to the pole of $\Pi$, which corresponds to the excitation spectrum of the superfluid Fermi gases.

In Fig.~\ref{fig2}(a), we show two types of excitation at the unitary limit, where both pair-breaking excitation and collective excitation are important. The pair-breaking excitation spectrum $\omega_{pb}$ corresponds to the poles of $\Pi_{pb}(\mathbf{q},z)$, namely $\omega_{pb}=E_++E_-$. It is a single-particle continuum, and its minimum $\omega_{\text{th}}(\mathbf{q})$ denotes the threshold energy to break a Cooper pair with center of mass momentum $\mathbf{q}$. The shaded region denotes that the imaginary part of the polarization differs from zero, and is referred to as the pair-breaking continuum. The collective spectrum $\omega_{\text{col}}(\mathbf{q})$ can be found by seeking the poles of $\Pi_{\text{cl}}(\mathbf{q},z)$, yielding  $I_{11}(\mathbf{q},\omega)I_{22}(\mathbf{q},\omega)-\omega^2I_{12}^2(\mathbf{q},\omega)=0$. The collective excitation spectrum exhibits characteristic linear energy-momentum behavior at small momentum $q$ as it is a sound mode, and it lies below the pair-breaking threshold. The behaviors of the imaginary part of the polarization function for three typical momenta $q/k_F=1.0$,$1.5$ and $2.0$ are shown in Fig.~\ref{fig2}(b). For $q/k_F=1.0$ and $1.5$, they have the same threshold energy $2\Delta_0$, below which $\text{Im}\Pi(q,\omega)$ vanishes, while for $q/k_F=2.0>4\mu_F^c$, the threshold energy is given by $\omega_{\text{th}}=\sqrt{(q^2/4-\mu^c)^2+\Delta_0^{2}}$. The magnitude of $\text{Im}\Pi(q,\omega)$ reaches maximum right after the threshold energy and decreases quickly with increasing energy.

The behaviors of the Bogoliubov spectrum $\omega(q)$ and the damping rate $\gamma(q)$ for three typical mass ratios $r_m=m_B/m_F$ are shown in the lower panel of Fig.~\ref{fig2}. At small momentum, the spectrum is phononlike with the sound velocity given by $c=\sqrt{(g_{B}+g_{BF}^2\Pi_0)n_B/m_B}$. For $r_m=1$, as shown in panel (c), there is a cusp in the spectrum resulting from the avoid-crossing of collective modes of the Fermi superfluid and the Bose superfluid. For sufficient large momentum $q$, both spectrums for $r_m=1$ and $r_m=1.5$ are entering the pair-breaking continuum, signifying that the Bogoliubov quasiparticle achieves finite lifetime due to damping effects. The damping occurs when the quasiparticle energy reaches the threshold energy $\omega_{\text{th}}$, swiftly reaches its maximum and decreases gradually for increasing momentum. Remarkably, there exists a critical mass ratio $r_m=2.7$, above which Bogoliugov excitations can achieve infinite lifetime with no damping. This special line of spectrum for $r_m=2.7$ intercepts the curve of the threshold energy at a critical momentum momentum $q_c$, and the damping vanishes for arbitrary momentum, as shown in panel (d).
\begin{figure}[t]
\includegraphics[width=8cm]{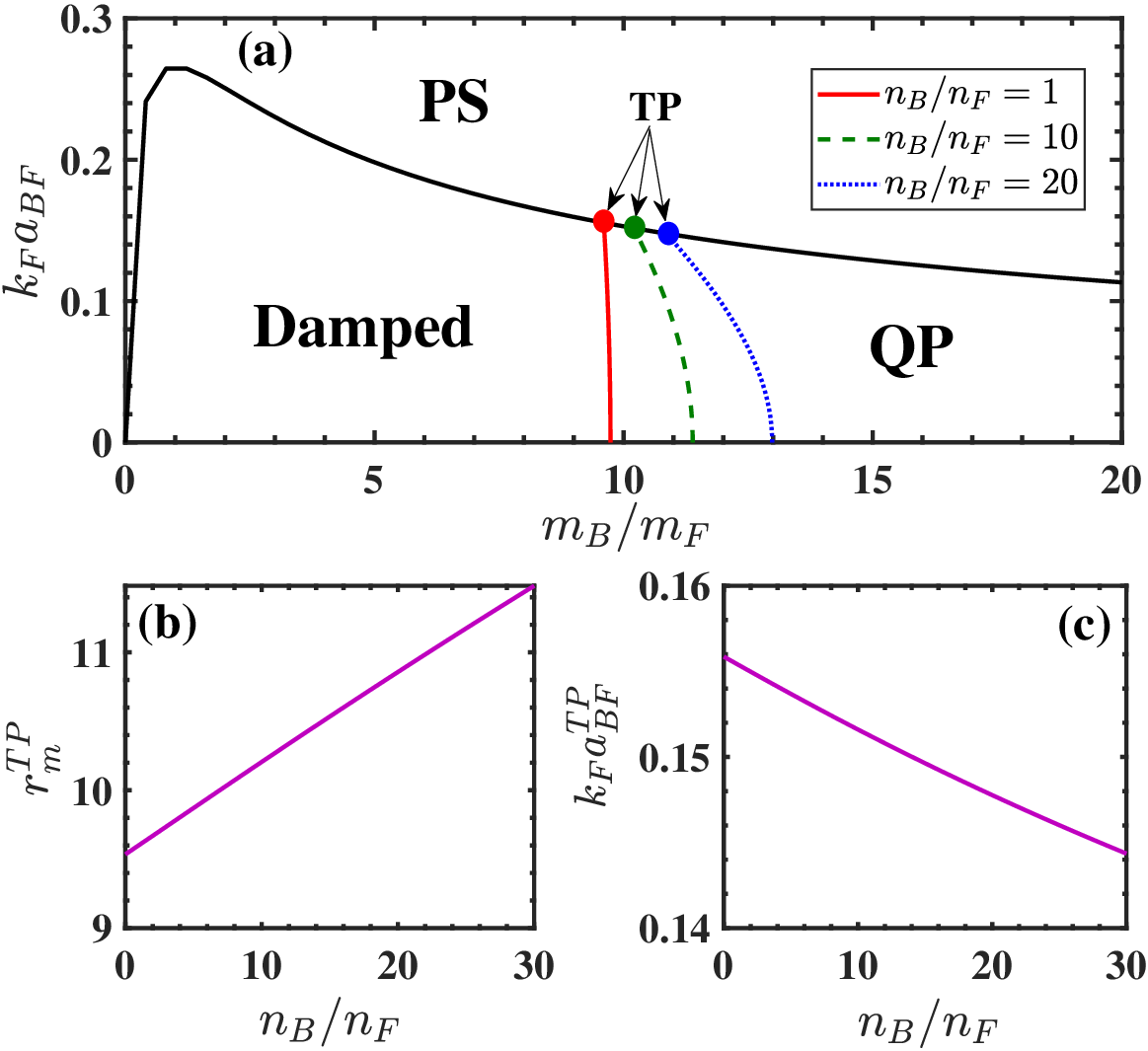}
\caption{(color online) Upper panel: Phase diagram spanned by boson-fermion mass ratio and coupling strength $k_Fa_{BF}$ at $1/k_Fa_{FF}=-1$ (BCS side). It has three regions: phase separation (PS), quasiparticle with infinite lifetime (QP) and damped region where quasiparticle has finite lifetime due to damping. For given boson-fermion density ratio, the three regions meet at a tricritical point (TP). Lower panel: The evolution of the tricritical point ($r_m^{TP}$, $k_Fa_{BF}^{TP}$) as a function of boson-fermion density ratio $n_B/n_F$.}
\label{fig3}
\end{figure}
\begin{figure}[t]
\includegraphics[width=8cm]{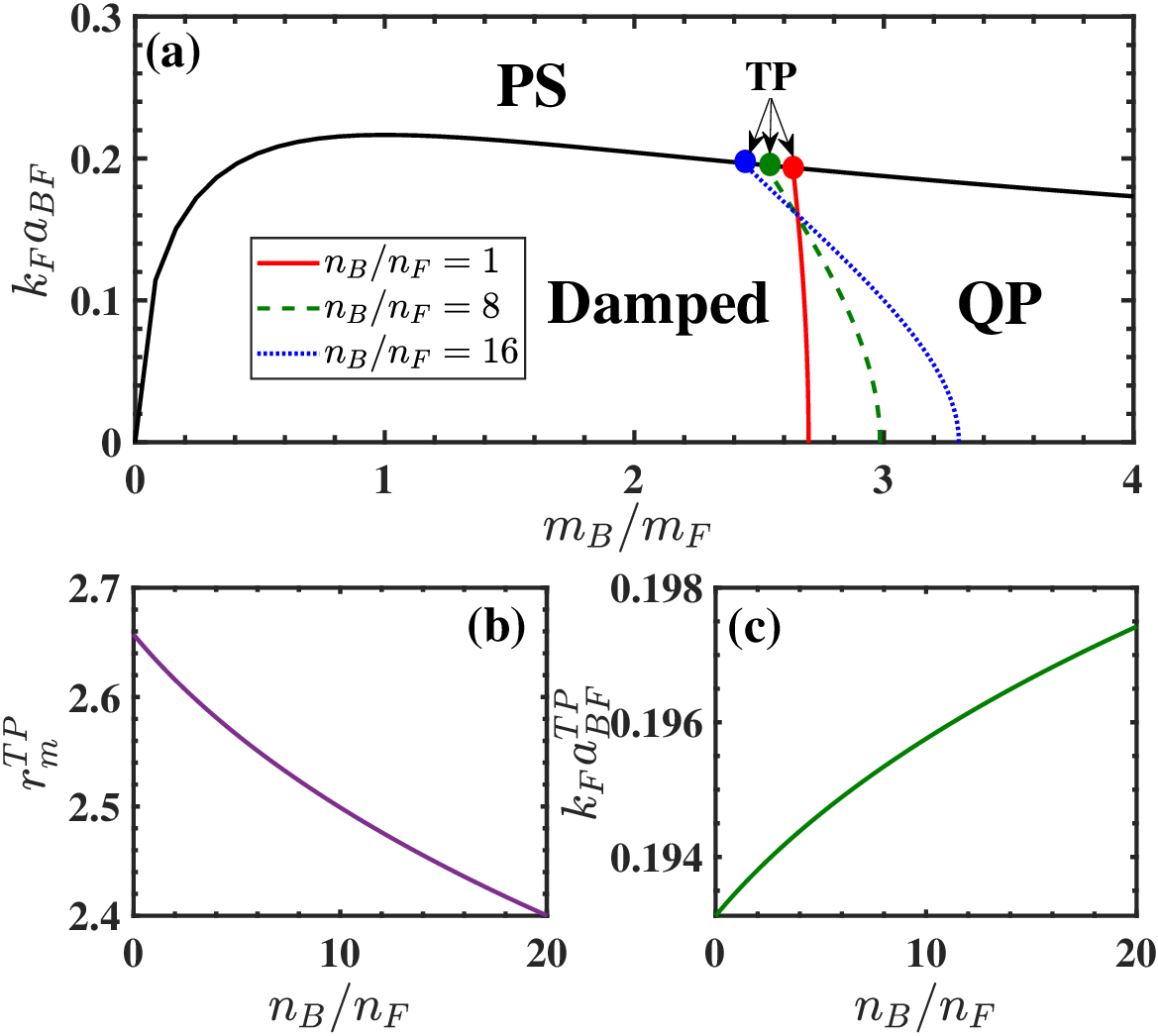}
\caption{(color online) Upper panel: Phase diagram spanned by mass ratio and boson-fermion coupling strength $k_Fa_{BF}$ at $1/k_Fa_{FF}=0$ (Unitarity). It has three regions: phase separation (PS), quasi-particle with infinite lifetime (QP) and damped region where quasiparticle has finite lifetime due to damping. For given boson-fermion density ratio, the three regions meet at a tricritical point (TP). Lower panel: The evolution of the tricritical point ($r_m^{TP}$, $k_Fa_{BF}^{TP}$) as a function of boson-fermion density ratio $n_B/n_F$.}
\label{fig4}
\end{figure}

We are now in position to construct a phase diagram for the system. The stability constraint marks the transition line between stable phase mixing and phase separation (PS) into fermions and bosons~\cite{MOL98,VIV00,ROT02,DAL16}, which remains the same for different number density ratio $n_B/n_F$, as shown in Fig.\ref{fig3} and Fig.\ref{fig4}, corresponding to $1/k_Fa=-1$ (BCS side) and $1/k_Fa=0$ (unitarity limit), respectively. In the stable phase mixing region, we can further classify it into regions accommodating quasiparticle excitations with and without damping, termed as damped and QP, respectively. To map out the phase boundary separating damped region and QP region, one needs to require that at the phase boundary the quasiparticle spectrum $\omega(q)$ is the tangent line to the threshold energy $\omega_{\text{th}}(q)$, which simultaneously determines both critical momentum $q_c$ and critical mass ratio $r_m$, illustrated previously in Fig.~\ref{fig2}(c).

At the BCS side with $1/k_Fa=-1$, as shown in the upper panel of Fig.~\ref{fig3}, the largest boson-fermion coupling strength $k_Fa_{FB}$ one can achieve to sustain a homogenous phase increases sharply, reaches a peak with $k_Fa_{FB}=0.27$ at $m_B/m_F=1$ and decreases slowly with increasing boson-fermion mass ratio $r_m$. As the boson-fermion density ratio $n_B/n_F$ increases, the regime of QP  diminishes, giving way to damped region. The tricritical point TP ($r_m^{TP}$, $k_Fa_{FB}^{TP}$) where the three phases meet can be tuned to move with density ratio $n_B/n_F$, as shown in the lower panel of Fig.~\ref{fig3}. The tricritical point moves toward high boson-fermion mass ratio and low boson-fermion coupling strength when $n_B/n_F$ increases.
\begin{figure}[t]
\centering
\includegraphics[width=8cm]{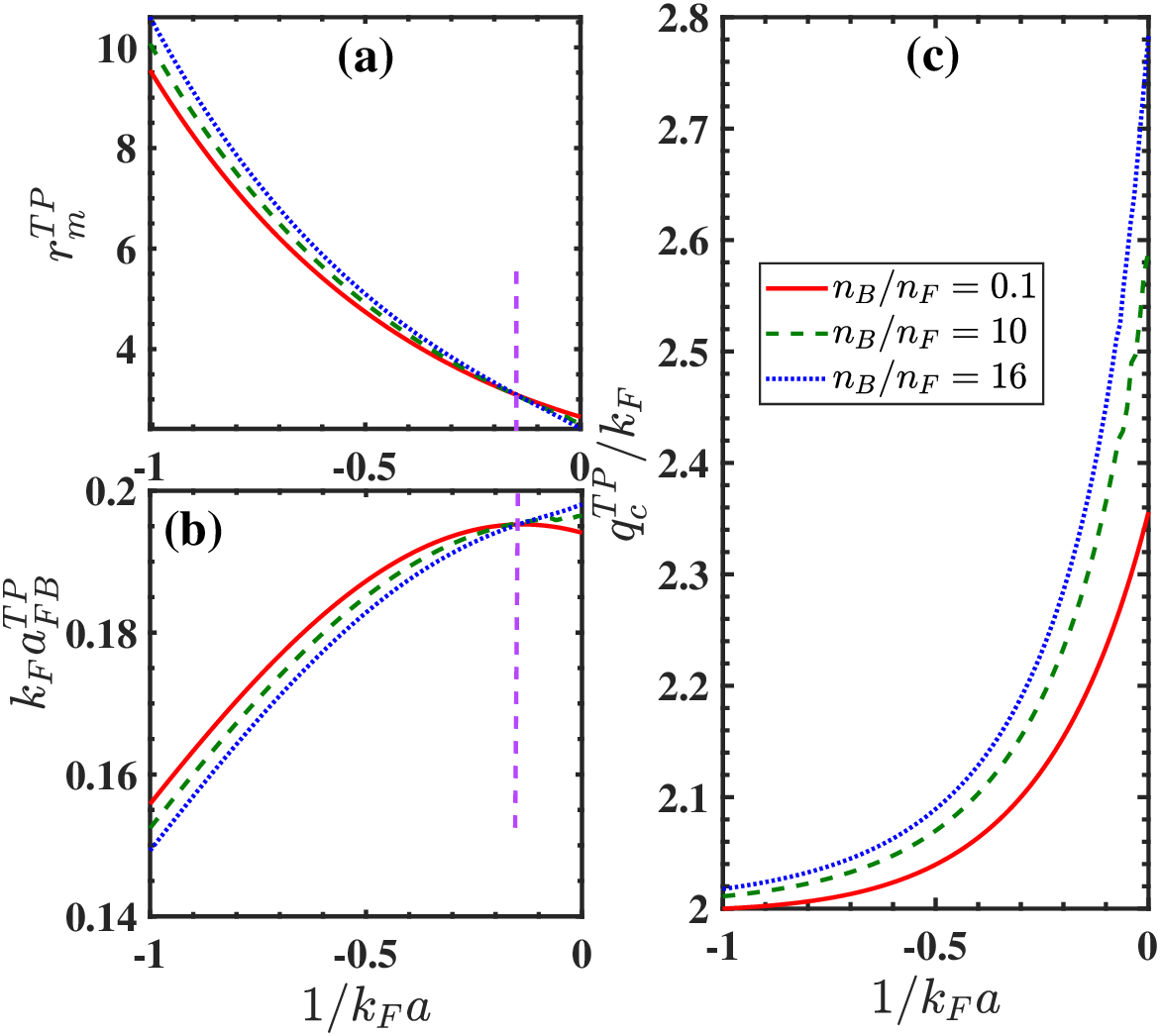}
\caption{(color online) Left panel: Evolution of the TP ($r_m^{TP}$,$k_Fa_{BF}^{TP}$) as a function of $1/k_Fa$. Right panel: Evolution of the critical momentum $q_c^{TP}$ at the TP as a function of $1/k_Fa$.}
\label{fig5}
\end{figure}

At the unitarity limit where $1/k_Fa=0$, as shown in upper panel of Fig.~\ref{fig4}, the phase boundary line between the phase mixing and phase separation varies smoothly with increasing boson-fermion mass ratio $m_B/m_F$. The phase diagram accommodates large portions of QP, as the boundary for boson-fermion mass ratio could reach $m_B/m_F=2.65$ when $n_B/n_F\rightarrow 0$. As $n_B/n_F$ increases, the tricritical point TP moves toward low mass ratio $r_m$ and high boson-fermion coupling strength $k_Fa_{BF}$, in stark contrast to the case of $1/k_Fa=-1$.

How does the interaction parameter $1/k_Fa$ control the motion of  the tricritical point becomes an interesting thing to investigate. This is shown in Fig.~\ref{fig5}. For all three typical density ratio $n_B/n_F=0.1$, $8$ and $16$, the critical mass ratio $r_m^{TP}$ decreases as one tunes up the BCS coupling strength $1/k_Fa$, as seen in panel (a). Conversely, the critical boson-fermion coupling strength $k_Fa_{BF}$ increases as one ramps up the BCS coupling strength $1/k_Fa$, as seen in panel (b). What is striking is that the behavior of TP as a function of density ratio $n_B/n_F$ shows reverse trend when it touches a critical BCS coupling strength roughly at $1/k_Fa=-0.16$. However, the critical momentum $q_c^{TP}$ follows the same trends for both BCS coupling strength and density ratio, as evident in panel (c).

\section{Conclusions}
In summary, we have investigated the superfluid mixtures of bosonic and fermionic atoms. By using the functional integral method to trace out the fermionic degrees of freedom, the effective action of the system shows that the induced interaction mediated by fermions between bosons are attractive interactions, it shows long-range behavior in BCS regime and gradually becomes short-ranged when it is driven toward BEC limit. By analyzing the Bogoliubov spectrum and the damping rate of bosonic superfluid, we have mapped out the phase diagram in the parameter space spanned by the boson-fermion mass ratio and the boson-fermion coupling strength, which shows that the stable phase mixing region can be further classified by damping of excitations, leading to a tricritical point in the phase diagram. A series of new features arising from fermion-mediated interactions have also been identified. The predicted damping rate can be probed experimentally via
two-phonon Bragg spectroscopy~\cite{DAV05}. Experimental verification of the predicted phase diagram will constitute an important step along the lines of searching for well-defined quasiparticle excitations in these systems. We hope that our work can add new excitement to the surging field of cold atom physics involving fermion-mediated interactions.
\section*{acknowledgments}
R.L. acknowledges funding from the NSFC under Grant No.12174055 and No.11674058, and by the Natural Science Foundation of Fujian under Grant No. 2020J01195. Lin Wen acknowledges the funding from the NSFC under Grant No. 12175027.
\begin{appendices}
\section{Derivation of the effective action}
\end{appendices}

\begin{thebibliography}{56}
\bibitem [{NOT()}]{NOTEOP}%
  \BibitemOpen
  \href@noop {} {\ }\bibinfo {note} {The superflid order parameter can be
  determined self-consistently as
  $\Delta(x)=-g_F\langle\psi_\downarrow(x)\psi_\uparrow(x)\rangle$ by imposing
  the saddle-point condition $\delta S/\delta \Delta(x)=0$}\BibitemShut
  {NoStop}%
\end{thebibliography}
%

\end{document}